%% file: main.tex
\documentclass{article}
\usepackage{ijcai24}

\usepackage{soul}
\usepackage[hidelinks]{hyperref}
\usepackage[utf8]{inputenc}
\usepackage[small]{caption}
\usepackage{amsthm}
\usepackage{algorithm}
\usepackage{algorithmic}
\usepackage[switch]{lineno}
\usepackage{graphicx}	
\usepackage{amsmath}	
\usepackage{amssymb}	
\usepackage{booktabs}
\usepackage{times}
\usepackage{microtype}
\usepackage{epsfig}
\usepackage[table,xcdraw,dvipsnames]{xcolor}
\usepackage{caption}
\usepackage{float}
\usepackage{placeins}
\usepackage{color, colortbl}
\usepackage{stfloats}
\usepackage{enumitem}
\usepackage{tabularx}
\usepackage{xstring}
\usepackage{multirow}
\usepackage{xspace}
\usepackage{url}
\usepackage{subcaption}
\usepackage{xcolor}
\usepackage[hang,flushmargin]{footmisc}
\usepackage{fp}

\DeclareMathOperator*{\argmin}{arg\,min}

\title{Slicer Networks}

\author{
Hang Zhang$^1$ \and
Renjiu Hu$^1$ \and
Xiang Chen$^2$ \and \\
Rongguang Wang $^3$ \and 
Dongdong Liu $^4$ \And
Gaolei Li $^5$
\\
\affiliations
$^1$Cornell University \qquad  $^2$Hunan University \qquad $^3$University of Pennsylvania\\
$^4$New York University \qquad $^5$Shanghai Jiao Tong University\\
}

\begin{document}

\maketitle

\input{docs/abstract}

\input{docs/intro}

\input{docs/related}
\input{docs/method}

\input{docs/results}
\input{docs/conclusion}

\appendix




\newpage

{
\small
\bibliographystyle{named}
\bibliography{ijcai24}
}
\end{document}

%% file: docs/abstract.tex
\begin{abstract}
In medical imaging, scans often reveal objects with varied contrasts but consistent internal intensities or textures. 
This characteristic enables the use of low-frequency approximations for tasks such as segmentation and deformation field estimation. 
Yet, integrating this concept into neural network architectures for medical image analysis remains underexplored.
In this paper, we propose the Slicer Network, a novel architecture designed to leverage these traits. 
Comprising an encoder utilizing models like vision transformers for feature extraction and a slicer employing a learnable bilateral grid, the Slicer Network strategically refines and upsamples feature maps via a splatting-blurring-slicing process. 
This introduces an edge-preserving low-frequency approximation for the network outcome, effectively enlarging the effective receptive field. 
The enhancement not only reduces computational complexity but also boosts overall performance.
Experiments across different medical imaging applications, including unsupervised and keypoints-based image registration and lesion segmentation, have verified the Slicer Network's improved accuracy and efficiency.
\end{abstract}

%% file: docs/intro.tex
\section{Introduction}
\label{sec:intro}

Recent developments in medical image analysis have benefited from the integration of Convolutional Neural Networks (ConvNets) \cite{he2016deep} and vision-based transformers \cite{liu2021swin,dosovitskiyimage}. 
These techniques have enhanced various tasks including lesion segmentation \cite{isensee2021nnu,zhang2023spatially2}, deformable image registration \cite{balakrishnan2019voxelmorph,zhang2023spatially}, and inverse reconstruction \cite{genzel2022near}. 
Two vital factors to their effectiveness are: 1) The access to large-scale datasets \cite{ridnik2021imagenet,sun2017revisiting} which are crucial for pretraining these backbone networks, and 2) The increased \textit{effective receptive field (ERF)} \cite{luo2016understanding} resulting from stacking deeper convolutional layers or using larger kernels \cite{ding2022scaling,liu2022convnet} in ConvNets and the integration of self-attention mechanisms in transformers \cite{vaswani2017attention,liu2021swin,dosovitskiyimage,wang2021pyramid}.

\begin{figure}[!t]
	\centering
	\subfloat[Unet decoder L1]{\includegraphics[width=.30\columnwidth]{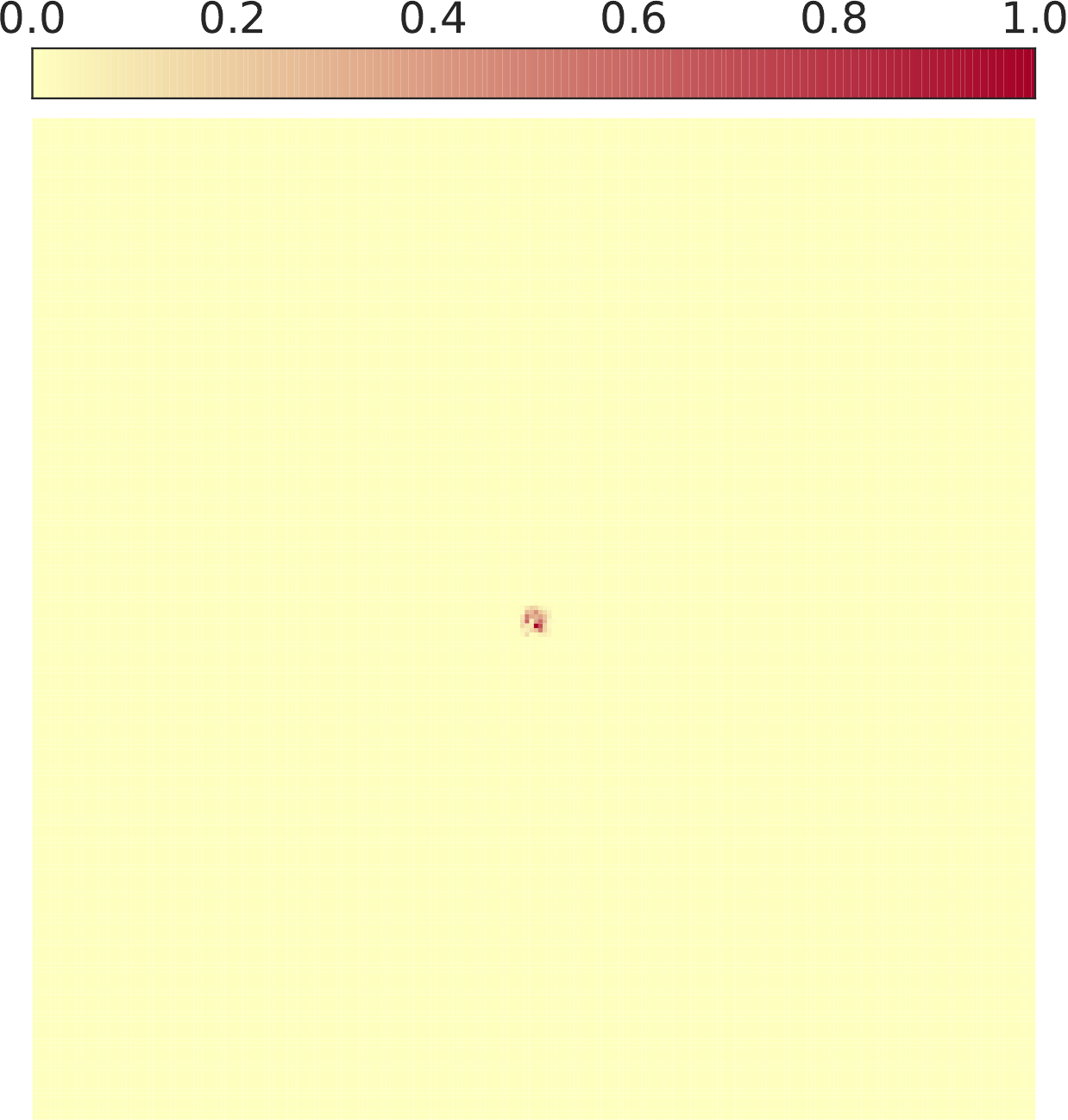} \label{fig:untrained_unet_erf_crop}}
    \subfloat[Swin-Unet]{\includegraphics[width=.30\columnwidth]{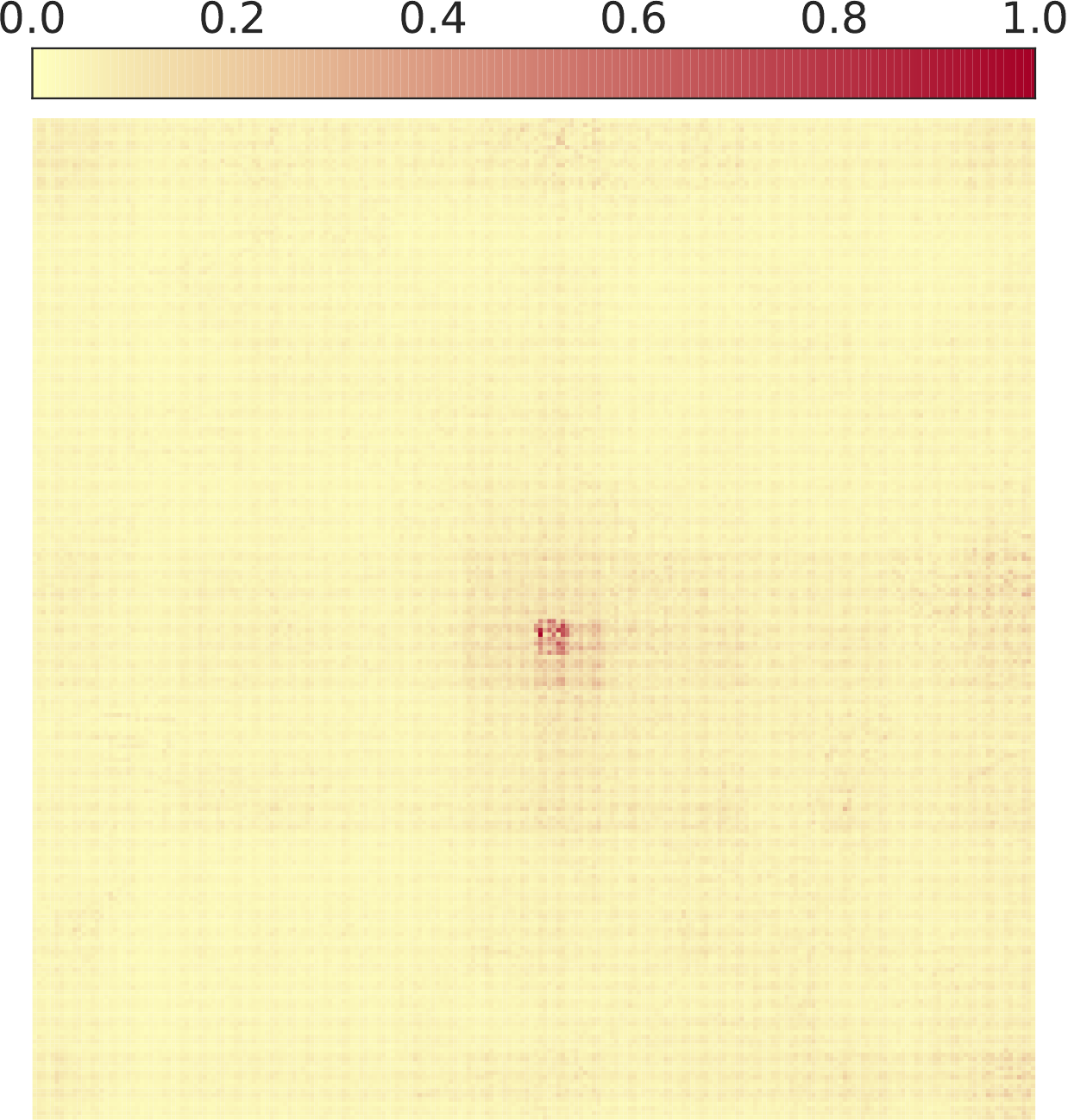} \label{fig:untrained_swinunet_erf_crop}}
    \subfloat[Swin-Slicer]{\includegraphics[width=.30\columnwidth]
    {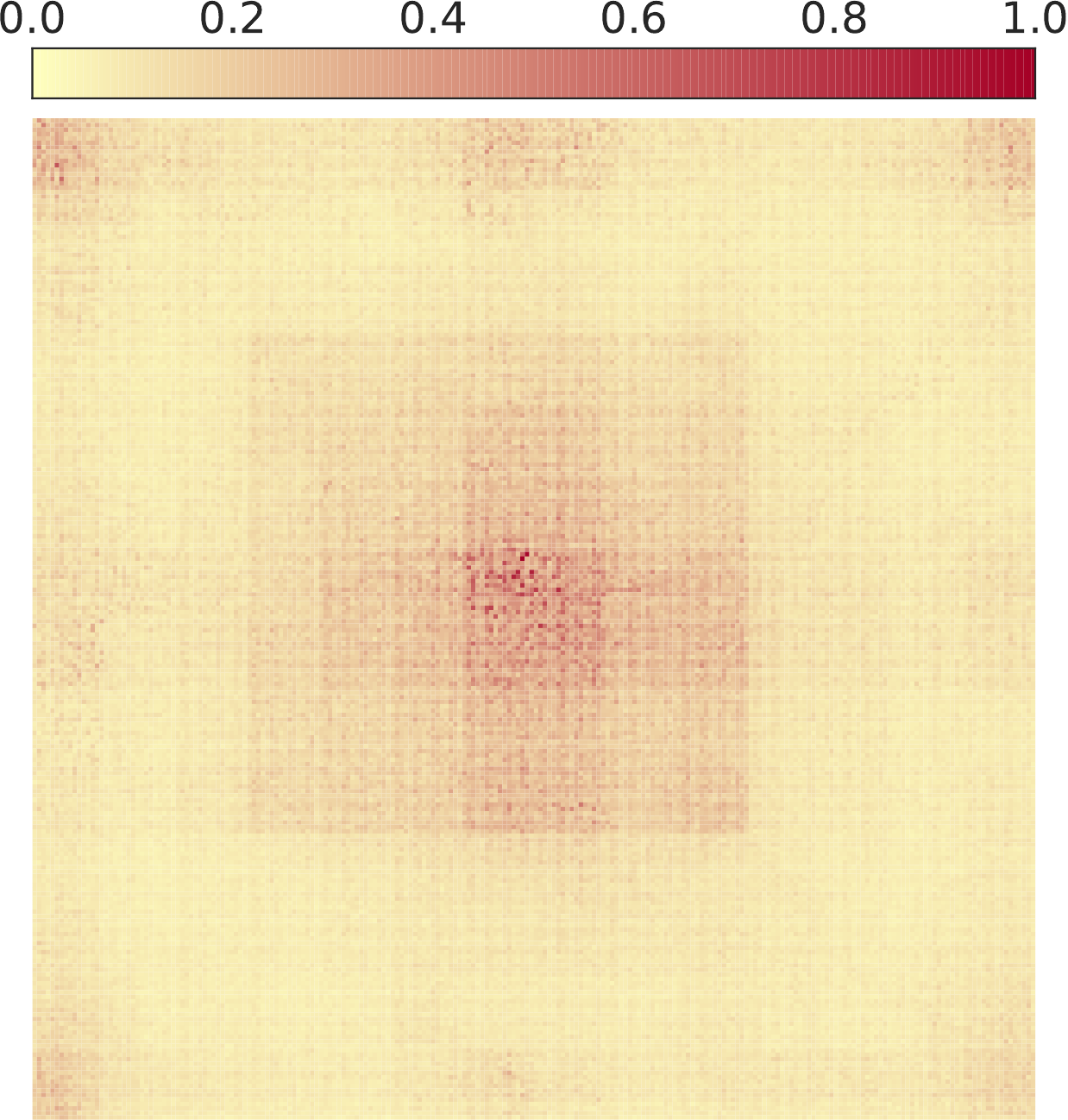} \label{fig:untrained_swinslicer_erf_crop}} \\
	\subfloat[Unet decoder L2]{\includegraphics[width=.30\columnwidth]
    {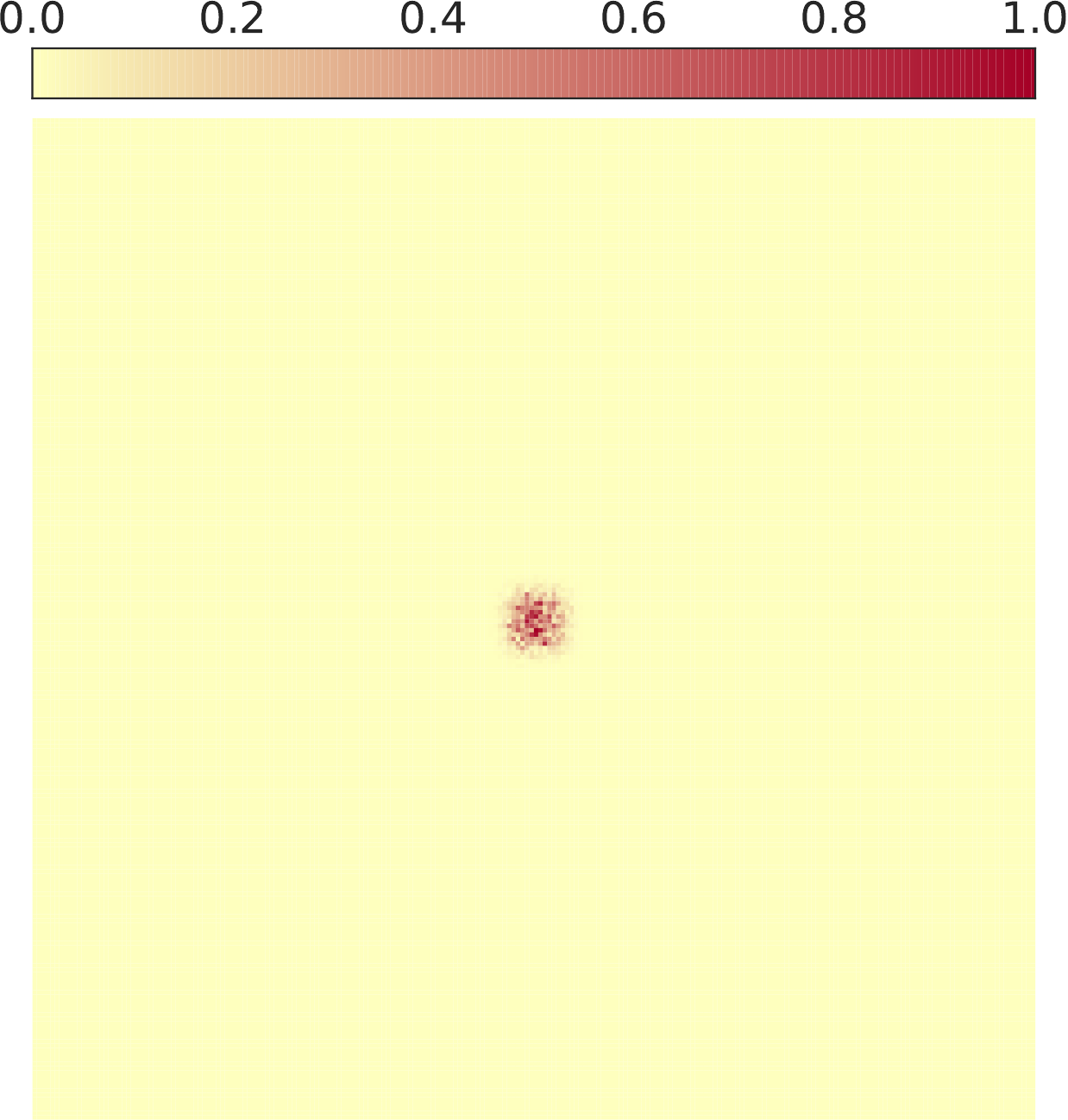} \label{fig:untrained_unet_erf_crop-2}}
    \subfloat[Unet decoder L3]{\includegraphics[width=.30\columnwidth]{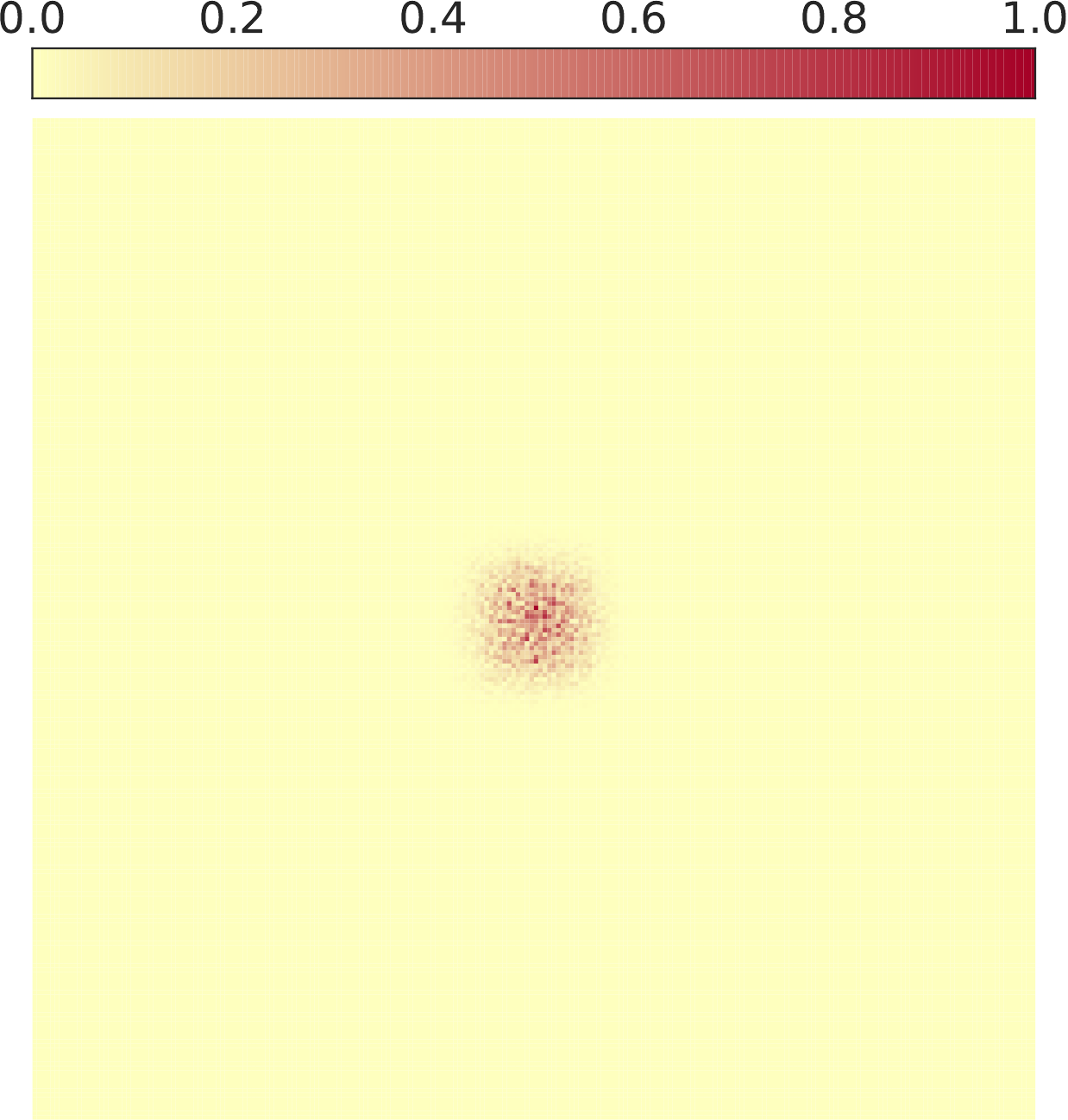} \label{fig:untrained_unet_erf_crop-3}}
    \subfloat[Unet decoder L4]{\includegraphics[width=.30\columnwidth]{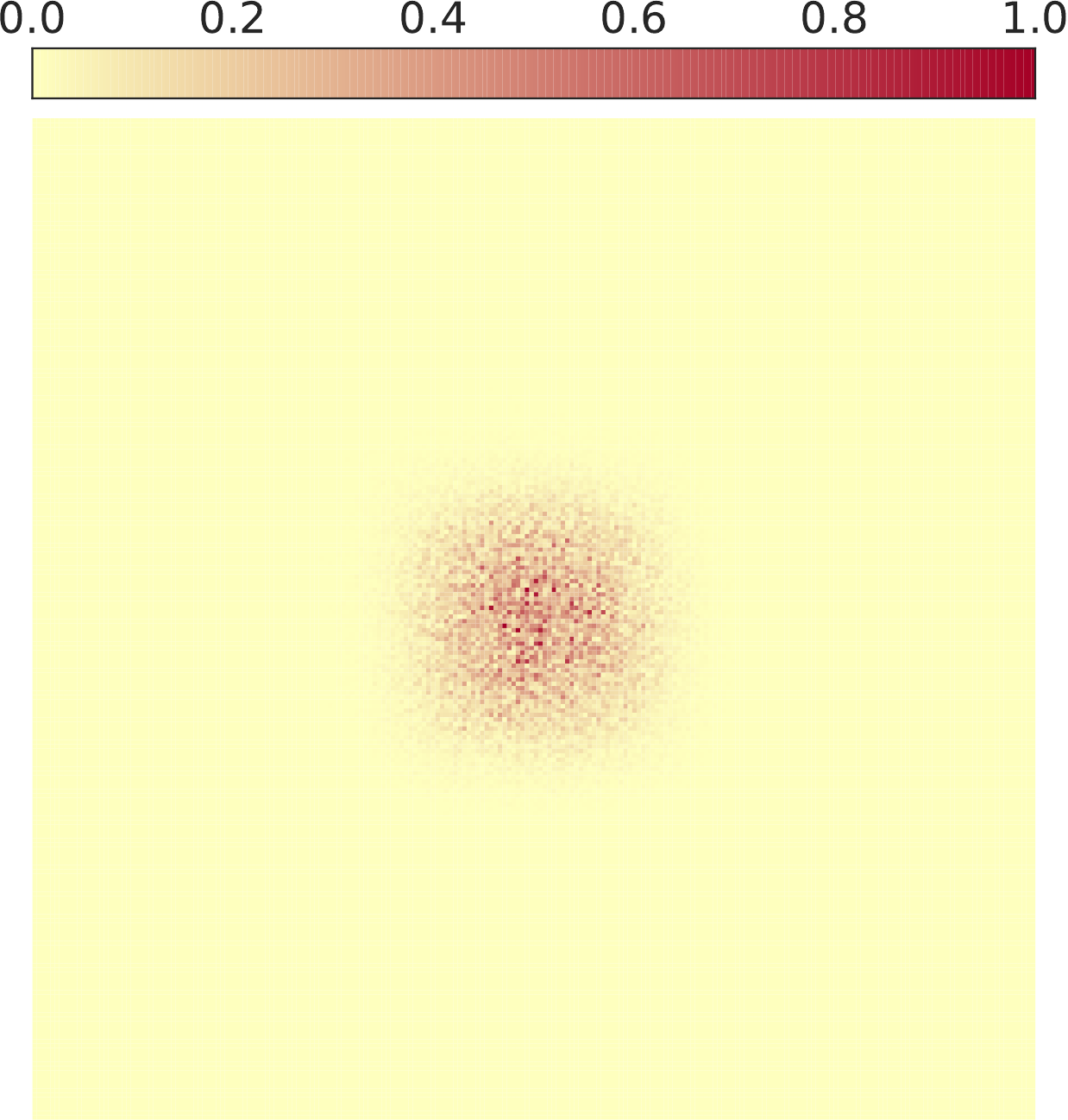} \label{fig:untrained_unet_erf_crop-4}}
 
    \caption{ 
        \textit{Effective Receptive Field (ERF)} visualizations \protect\cite{luo2016understanding} across architectures: U-Net (a, d, g, j), Swin-Unet (b), Swin-Slicer (c). 
        Darker and more widely spread regions indicate larger ERFs. 
        Swin-Unet and Swin-Slicer feature maps are presented pre-softmax, while Unet utilizes decoder feature maps, with L4 to L1 showing increased spatial sizes via upsampling. 
        Our slicer network enhances the ERF compared to standard Unet layers.
    }
    \vspace{-1ex}
	\label{fig:erf_examples}
\end{figure}

Building upon these foundational backbones, a variety of U-shaped encoder-decoder networks \cite{ronneberger2015u} have been developed, as detailed in works like \cite{cao2022swin,chen2021transunet,rahman2023medical,tang2022self,lee20223d}. 
These networks typically combine an encoder with cascaded decoders, connected via shortcuts. 
This design, while proficient in detailing fine attributes, tends to diminish the ERF, which may result in reduced contextual understanding (refer to Fig. \ref{fig:erf_examples}). 
This study aims to tackle this limitation from a network architecture perspective, seeking to enhance both the effectiveness and efficiency of certain medical image analysis tasks.

\begin{figure}[!t]
	\centering
    \subfloat[Splatting]{\includegraphics[width=.49\columnwidth]{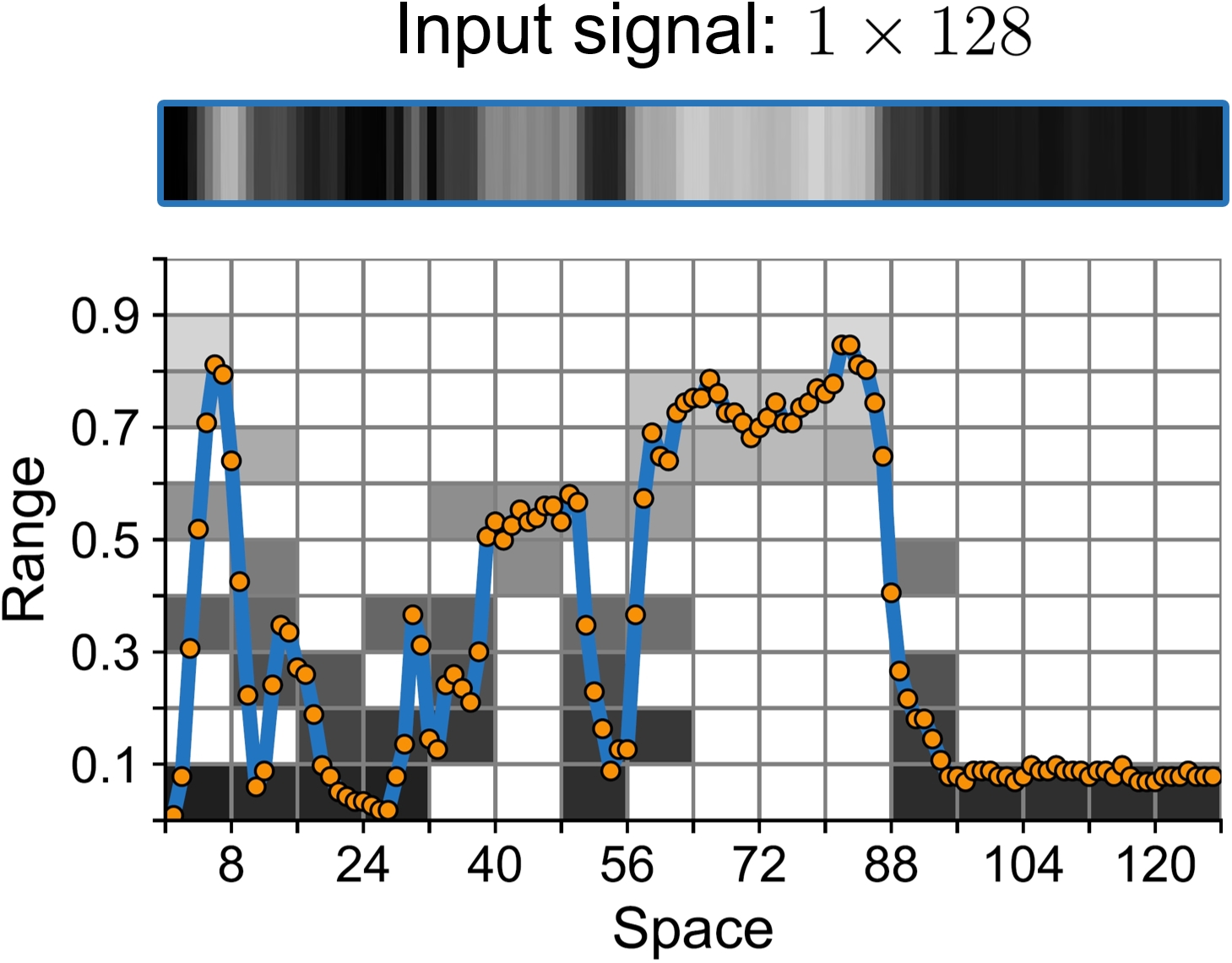} \label{fig:bg_splatting}} 
    \subfloat[Blurring and slicing]{\includegraphics[width=.49\columnwidth]{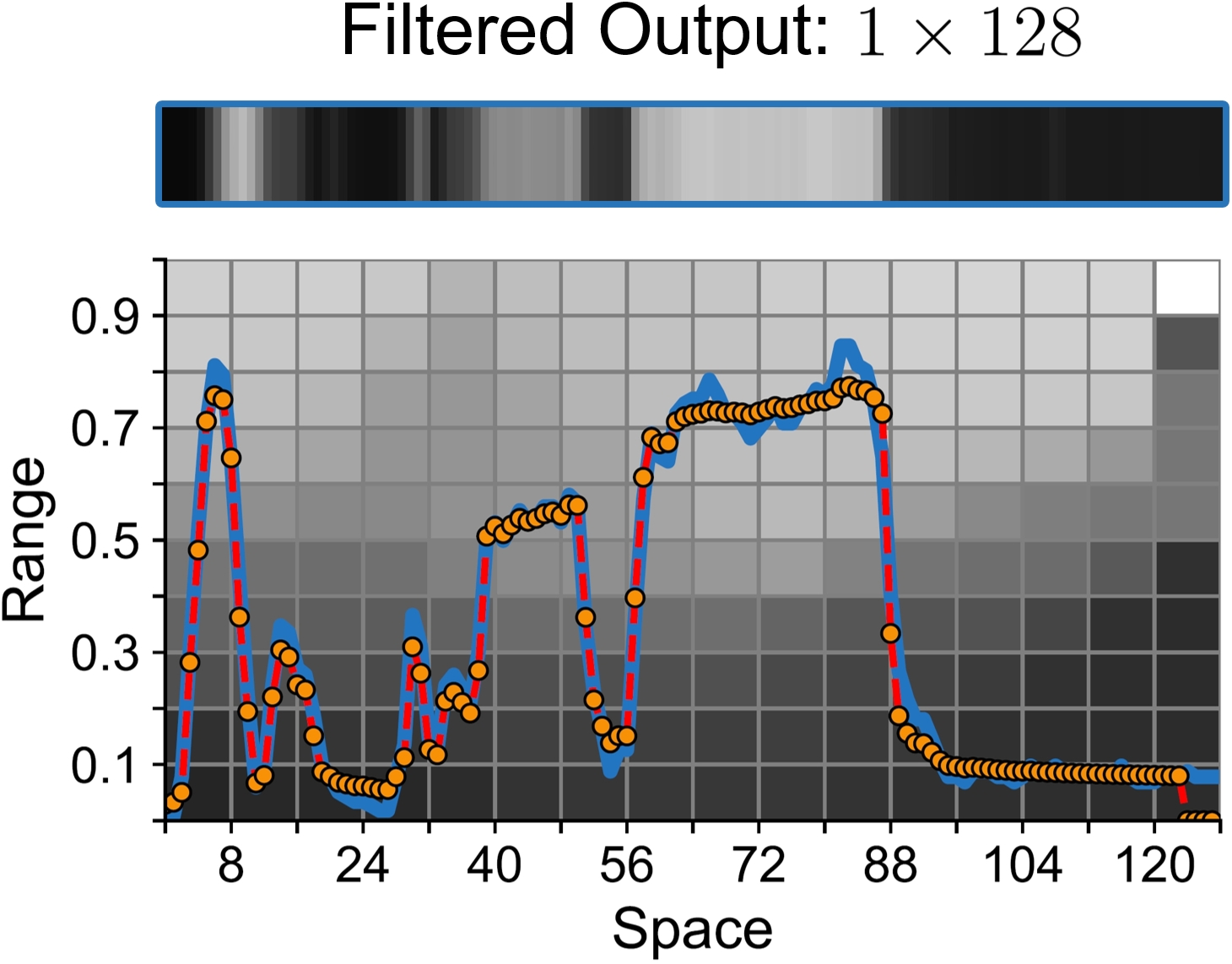} \label{fig:bg_blurring}}
    \caption{ 
        Visual illustration of the traditional tri-phase bilateral grid process comprising splatting, blurring, and slicing on a 1D signal.
        \textbf{Splatting}: Employs a nearest sampling kernel to transform image signals, normalized to (0,1) from a dimension of $1 \times 128$ into a $16 \times 10$ sparse spatial-range grid using sampling rates $s_s=8$ and $s_r=0.1$.
        \textbf{Blurring}: Applies Gaussian filters with $\sigma=1$ to both spatial and range dimensions, corresponding to $\sigma_s=\sigma \times s_s$ and $\sigma_r=\sigma \times s_r$ in the initial image space.
        \textbf{Slicing}: Utilizes a bilinear sampling kernel, normalized via homogeneous coordinates.
        (a) Depicts the bilateral grid aligned with the input 1D signal, with orange dots showing projected intensities linked by a blue line.
        (b) Exhibits the Gaussian-blurred grid and filtered output 1D signal; orange dots are sliced intensities, contrasted against the original blue line using a dashed red line.
    }
     \vspace{-1ex}
	\label{fig:bg_explain}
\end{figure}

Imaging modalities such as Magnetic Resonance Imaging (MRI), Computed Tomography (CT) and dermoscopy produce numerous images for inspection on specific targets like cardiac ventricles or skin lesions.
These images usually require registration or segmentation before detailed analysis.
Many exhibit consistent internal contrasts and textures within objects and clear variations across boundaries.
This characteristic imparts a \textit{piece-wise smooth} quality to both the images and their outputs, including segmentation masks and deformation fields. 
With careful preservation of object boundaries, accurate low-frequency approximations could be effective without losing critical details.
For example, the bilateral grid \cite{paris2006fast,chen2007real} and its subsequent variants \cite{adams2009gaussian,gastal2012adaptive,gharbi2017deep} have been widely used to accelerate megapixel image processing tasks via low-frequency approximations in high-dimensional grid space.
Fig. \ref{fig:bg_explain} illustrates how image signals are projected in this context.
Addressing the outlined challenges and characteristics leads to an assumption and two pivotal questions:
\textbf{Assumption:} Medical imaging tasks with \textit{piece-wise smooth} inputs and outputs lend themselves to effective low-frequency feature approximation.
\textbf{Q1:} How can we enlarge the network's ERF without losing critical details?
\textbf{Q2:} Is it possible to reduce computational demands for resource-limited clinical settings?

To tackle the questions within the assumption, we introduce the Slicer Network, a novel architecture that enhances the ERF and boosts performance of medical application, providing an alternative alongside the traditional encoder-decoder networks.
This framework is composed of two components: an encoder, adopted from established backbone architectures like Residual Network (ResNet) \cite{he2016deep} and Vision Transformers (ViTs) \cite{dosovitskiyimage}, and a slicer network with a learnable bilateral grid.
The encoder creates feature maps that abstract and compress input images into low-resolution representations; 
the slicer then upsamples these feature maps, using a guidance map derived from raw images, to produce outputs that maintain object boundary integrity. 
The Slicer Network is similar to a combination of bilateral grid \cite{paris2006fast,chen2007real} and joint bilateral upsampling \cite{kopf2007joint}, but it operates within a differentiable neural network context.

In our study, we evaluate our method across three medical imaging tasks: unsupervised cardiac MRI registration, keypoints-based lung CT registration, and dermoscopy-based skin lesion segmentation. 
Key findings include:
\begin{itemize}
\item The Slicer Network effectively expands the ERF while preserving boundary details, demonstrating superior performance in cardiac registration and skin lesion segmentation compared to other leading methods.
\item The integration of a learnable bilateral grid enables zero-shot learning for keypoints-based lung CT registration, enhancing performance beyond conventional methods.
\item In applications with \textit{piece-wise smooth} inputs and outputs, like cardiac registration, our approach can significantly reduce computational demands by at least 80\% while maintaining similar accuracy.
\end{itemize}

%% file: docs/related.tex
\section{Related Work}
\label{sec:related}

\subsection{Bilateral Gird and High-Dimensional Filtering}

The original bilateral filter, as proposed by \cite{smith1997susan} and \cite{tomasi1998bilateral}, enhances image quality by replacing each pixel with a weighted average of neighboring pixels. 
These weights are determined by both the spatial proximity within the image plane (\textit{spatial domain} $\mathcal{S}$) and the similarity in intensity values (\textit{range domain} $\mathcal{R}$). 
While this technique is effective at edge-preserving image manipulation, its native form suffers from slow processing speeds.
To address this, advancements such as the bilateral grid \cite{paris2006fast,chen2007real}, Gaussian KD-Trees \cite{adams2009gaussian} and adaptive manifolds \cite{gastal2012adaptive} have emerged. 
These techniques accelerate the bilateral filtering process by projecting original signals into a smaller yet higher-dimensional space, enabling real-time performance.
Subsequently, such techniques have been adapted into neural networks for tasks like scene-dependent image transformation \cite{gharbi2017deep} and stereo matching \cite{xu2021bilateral}. 
However, their usage in tasks such as image segmentation and image registration in the medical domain is limited. 
Innovations like bilateral neural networks \cite{jampani2016learning} and the fast bilateral solver \cite{barron2016fast} have sought to harness the potential of neural networks for these complex tasks, yet their applicability is still confined.

\subsection{Learning with Differentiable Transformations}
Grid Sampling (GS) \cite{jaderberg2015spatial} has empowered neural networks with transformation capabilities via differentiable interpolation. 
However, it is not suitable for transformations involving integration or summing, such as the Hough \cite{ballard1981generalizing,lin2020deep,zhao2021deep} and Radon \cite{helgason1999radon} transforms and bilateral grid creation \cite{paris2006fast,chen2007real}.
\cite{zhang2023deda} introduces the Deep Directed Accumulator (DeDA) to address this limitation, establishing a complementary relationship with GS. 
GS "pulls" values to each cell in the target feature map from the source, whereas DeDA "pushes" values from each cell in the source to the target feature map. 
Essentially, GS samples, while DeDA accumulates values from the source.
When applied with proper geometric parametrization, the integration transformation highlights key geometric structures like lines \cite{lin2020deep,zhao2021deep} and circles \cite{zhang2023deda} in new representations. 
Combined, DeDA and GS complete a forward-backward cycle in dense mapping tasks. 
Within our Slicer Network, DeDA constructs a differentiable bilateral grid, and GS performs slicing from this grid.

%% file: docs/method.tex
\section{Method}
\label{sec:method}
In this section, we present the Slicer Network, starting with the traditional bilateral grid basics. 
We then explore the slicer network which integrates a differentiable cross-bilateral grid in neural networks using DeDA and GS. 
For simplicity and without loss of generality, we illustrate our framework using a 2-dimensional spatial domain.

\subsection{Preliminaries}
We start this section with a brief review of the process of implementing cross-bilateral filtering via a bilateral grid \cite{paris2006fast,chen2007real} (ref to Fig. \ref{fig:bg_explain} for an example). 
Consider a guidance image $\mathbf{G}^r \in \mathbb{R}^{h \times w}$ normalized to the range $(0,1)$, and an input image $\mathbf{I} \in \mathbb{R}^{h \times w}$. 
Let $s_s$ and $s_r$ denote the sampling rates in the spatial domain $\mathcal{S}$ and the range domain $\mathcal{R}$, respectively. 
We can establish a bilateral grid $\mathbf{\Gamma} \in \mathbb{R}^{[\frac{h}{s_s}] \times [\frac{w}{s_s}] \times [\frac{1}{s_r}]}$. 
This grid is initially set to zero and then updated by accumulating homogeneous coordinates:
\begin{equation}
\mathbf{\Gamma}\left(\left[\frac{x}{s_s}\right], \left[\frac{y}{s_s}\right], \left[\frac{\mathbf{G}(x, y)}{s_r}\right]\right) \mathrel{+}= (\mathbf{I}(x, y), 1),
\label{eq:splatting_bg}
\end{equation}
where $x,y$ are coordinates, and $\left[\cdot\right]$ is the rounding operation.

The process described in Eq.~\eqref{eq:splatting_bg} is known as splatting, wherein image signals are projected onto the higher-dimensional spatial-range domain $\mathcal{S} \times \mathcal{R}$. 
Any function $f$, including neural networks that inputs the constructed grid $\mathbf{\Gamma}$, can be utilized to manipulate the grid, resulting in $\tilde{\mathbf{\Gamma}} = f(\mathbf{\Gamma})$. 
For convolution operations, filtering within $\mathcal{S} \times \mathcal{R}$ respects locality both in the spatial and range domains. 
Subsequently, slicing is executed to formulate a new image by accessing the grid at locations $\left(\frac{x}{s_s}, \frac{y}{s_s}, \frac{\mathbf{G}(x, y)}{s_r}\right)$ through linear interpolation.

\begin{figure}[!b]
    \centering
    \includegraphics[width=0.85\columnwidth]{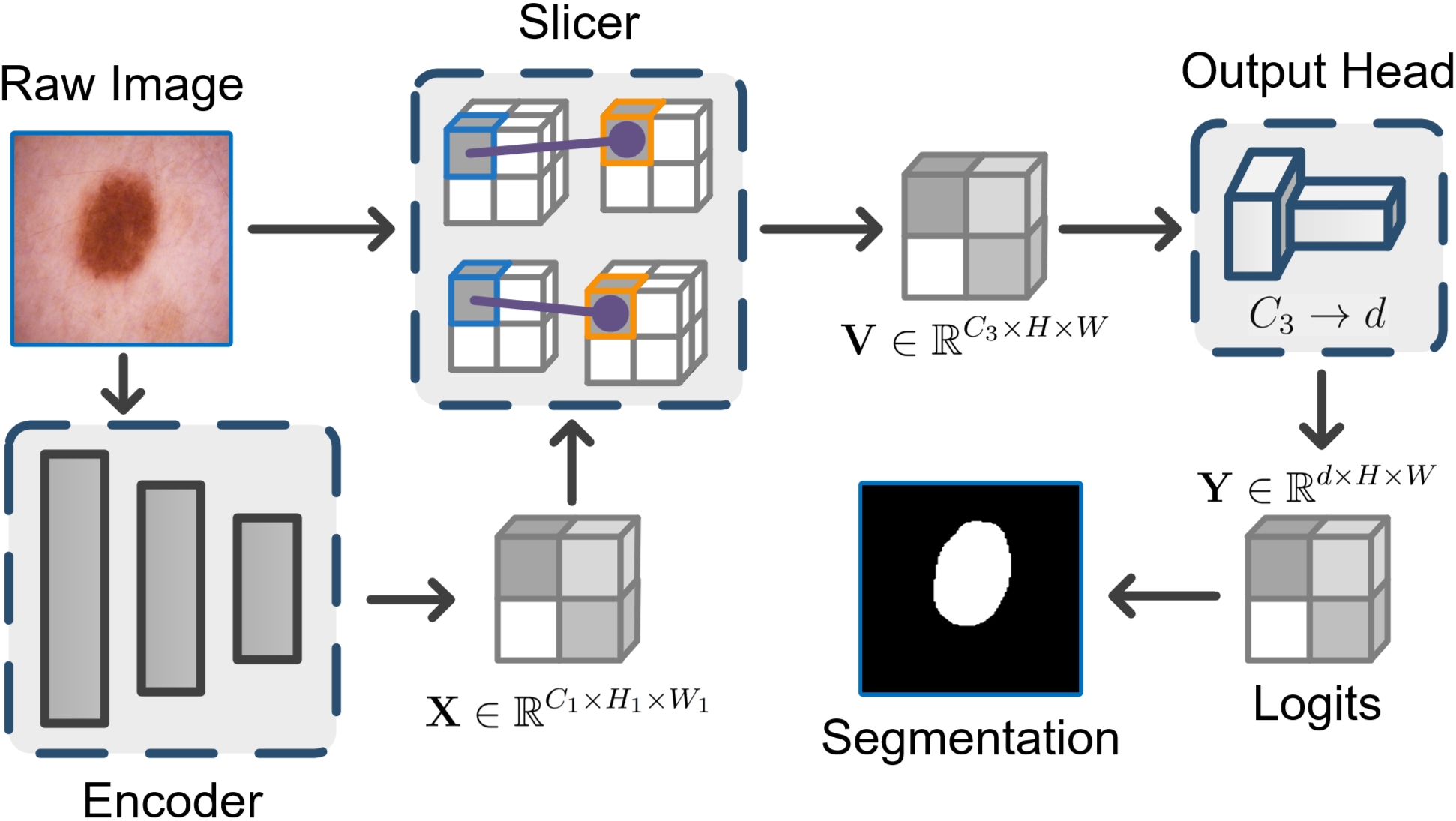}
    \caption{
        Overview of the Slicer Network framework. 
        The raw image generates a guidance map and an encoder-derived feature map, approximating low-frequency segmentation context. 
        The slicer then refines and upsamples this map, leading to the final segmentation logits via an output head.
    }
     \vspace{-1ex}
    \label{fig:encoder_slicer_framework}
\end{figure}

\subsection{Differentiable Splatting and Slicing}
\label{sec:splatting_slicing}
Adopting notations from prior research \cite{jaderberg2015spatial,zhang2023deda}, we formulate the splatting and slicing operations as outlined below. 
When dealing with multi-channel inputs, the transformation process is independently applied to each channel. 
Our discussion primarily focuses on the spatial dimensions of these components for clarity. 
Although we demonstrate using 2D images and 3D grids, adapting this to 3D volumes and 4D grids is straightforward.

\vspace{1ex}
\noindent\textbf{Splatting.} 
Given a source feature map $\mathbf{U} \in \mathbb{R}^{H \times W}$ and a target $\mathbf{\Gamma} \in \mathbb{R}^{H' \times W' \times R}$, with $H,W$ as input and $H',W'$ as output spatial dimensions, and $R$ as the range dimension size. 
A sampling grid $\mathbf{G} \in \mathbb{R}^{H \times W \times 3}=(\mathbf{G}^x, \mathbf{G}^y,\mathbf{G}^r)$, with $\mathbf{G}^x,\mathbf{G}^y$ as mesh grids and $\mathbf{G}^r$ as the guidance map, along with a kernel function $\mathcal{K}()$, are used. 
The value in the bilteral grid $\mathbf{\Gamma}$ at cell $(i,j,k)$ is defined as:
\begin{equation}
    \mathbf{\Gamma}_{ijk} = \sum_{(n,m)}^{\mathcal{S}} \mathbf{U}_{nm}\mathcal{K}(\mathbf{G}_{nm}^x,i)\mathcal{K}(\mathbf{G}_{nm}^y,j)\mathcal{K}(\mathbf{G}_{nm}^r,k),
    \label{eq:splatting}
\end{equation}
where $\mathcal{S}$ denotes the spatial domain, and the kernel function $\mathcal{K}()$ can be any predefined kernel, like the linear sampling kernel $\mathcal{K}(p,q)=\text{max}(0,1-|p-q|)$. 
Eq.~\eqref{eq:splatting} is a tensor mapping function by $\mathcal{D}:(\mathbf{U},\mathbf{G};\mathcal{K})\mapsto \mathbf{\Gamma}$.
with a linear kernel, both $\mathbf{G}$ and $\mathbf{U}$ backpropagate gradients,unlike the nearest kernel where only $\mathbf{U}$ does.
Different from the original bilateral grid which uses nearest kernel for grid creation, we use a linear kernel and found it to be more effective.

\vspace{1ex}
\noindent\textbf{Slicing.}
Slicing, the symmetric operation to splatting, is a crucial step that generates the piece-wise smooth output. 
For the 'blurred' bilateral grid $\tilde{\mathbf{\Gamma}} \in \mathbb{R}^{H' \times W' \times R}$, we determine the value of a specific cell $(n,m)$ in the sliced feature map $\mathbf{V} \in \mathbb{R}^{H\times W}$ as following:
\begin{equation}
    \mathbf{V}_{nm} = \sum_{(i,j,k)}^{\mathcal{S}\times \mathcal{R}} \tilde{\mathbf{\Gamma}}_{ijk}\mathcal{K}(\mathbf{G}_{nm}^x,i)\mathcal{K}(\mathbf{G}_{nm}^y,j)\mathcal{K}(\mathbf{G}_{nm}^r,k),
\label{eq:slicing}
\end{equation}
Here, $\mathcal{R}$ denotes the range domain, with other parameters as in Eq. \eqref{eq:splatting}.
The main distinction between Eq. \eqref{eq:splatting} (splatting) and Eq. \eqref{eq:slicing} (slicing) is data flow direction. 
In splatting, each cell $\mathbf{\Gamma}_{ijk}$ aggregates values from the source feature map that correspond to its location. 
Conversely, in slicing, each cell $\mathbf{V}_{nm}$ identifies its corresponding location in the source feature map and retrieves the value from there.
This grid process enables edge-aware functionality. 
Gradient details for both equations are in \cite{jaderberg2015spatial,zhang2023deda}.

\vspace{1ex}
\noindent\textbf{Homogeneous Coordinates.}
For edge-preserving filtering in the bilteral grid, tracking the pixel count or weight per grid cell is essential.
We store homogeneous coordinates $(\mathbf{\Gamma}_{ijk}\mathbf{W}_{ijk},\mathbf{W}_{ijk})$ during grid creation. 
Here, $\mathbf{W}$ is derived from Eq.~\eqref{eq:splatting} $\mathbf{W}=\mathcal{D}(\mathbf{J},\mathbf{G};\mathcal{K})$, with $\mathbf{J}$ as a tensor of ones. 
This facilitates weighted average calculations and normalizations.
Homogeneous coordinate $\mathbf{W}$ indicates the importance of its corresponding data $\mathbf{V}$.
In practice, $\mathbf{W}$ is obtained as an extra output channel, corresponding to $\mathbf{J}$ serving an additional input channel, alongside the image data.

\begin{figure}[!t]
    \centering
    \includegraphics[width=1.0\columnwidth]{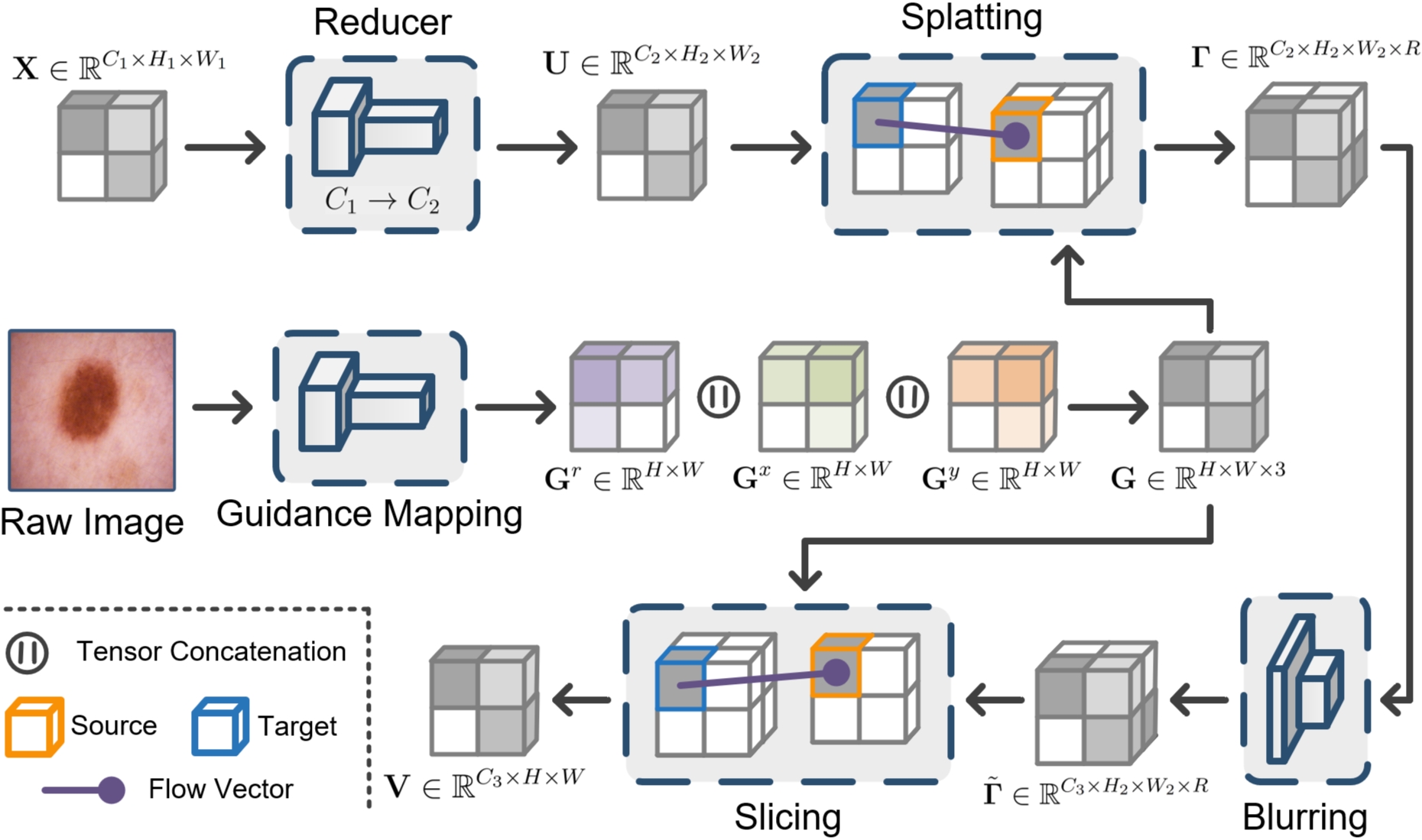}
    \caption{
        Visual illustration of the Slicer. 
        Details on the splatting and slicing processes are discussed in Section \ref{sec:splatting_slicing}. 
        The network's reducer, guidance mapping, and blurring components consist of convolutional layers.
    }
     \vspace{-1ex}
    \label{fig:slicer_network}
\end{figure}

\subsection{The Slicer Network}

Modern neural networks typically employ an encoder-decoder architecture such as Unet \cite{ronneberger2015u} for dense mapping tasks.
However, their \textit{effective receptive field (ERF)} \cite{luo2016understanding} is Gaussian and covers only part of the theoretical receptive field. 
Consecutive up-sampling layers may maintain the theoretical field but reduce the ERF, as shown in Fig. \ref{fig:erf_examples}. 
Upsampling directly from the encoder output preserves the ERF but may result in loss of details.
The Slicer Network, featuring an encoder and an edge-preserving slicer (see Fig. \ref{fig:encoder_slicer_framework} for a visual depiction), addresses this issue by allowing direct upsampling from the encoder output, retaining the ERF without losing details.

\vspace{1ex}
\noindent\textbf{The Slicer Framework.}
Central to the Slicer Network, as shown in Fig. \ref{fig:slicer_network}, is the slicer. 
The encoder output $\mathbf{X} \in \mathbb{R}^{C_1\times H_1 \times W_1}$ is reduced to $\mathbf{U} \in \mathbb{R}^{C_2\times H_2\times W_2}$, forming a low-frequency approximation for segmentation or deformation. 
The guidance map $\mathbf{G}^r$, combined with mesh grids, creates the sampling grid $\mathbf{G}$.
The feature map $\mathbf{U}$ is then mapped onto a bilateral grid $\mathbf{\Gamma} \in \mathbb{R}^{C_2 \times H_2 \times W_2 \times R}$, which undergoes a 'blurring' process via learnable convolution layers to produce the refined grid $\tilde{\mathbf{\Gamma}} \in \mathbb{R}^{C_3 \times H_2 \times W_2 \times R}$. 
Slicing this grid results in the feature map $\mathbf{V} \in \mathbb{R}^{C_3\times H \times W}$, matching the original spatial dimensions.

\vspace{1ex}
\noindent\textbf{Guidance Map Generator.}
The guidance image $\mathbf{G}^r$ represents the range domain in our bilateral grid. 
Traditional methods use luminance or RGB channels \cite{paris2006fast,chen2007real}, but learned guidance maps \cite{gharbi2017deep} are more effective. 
Our generator has two convolutional layers: a ReLU-activated first layer and a Sigmoid-activated second layer, producing a single-channel tensor, valued within $[0,1]$, with spatial dimensions matching the input.
This guidance map $\mathbf{G}^r$, concatenated with mesh grids $\mathbf{G}^x$ and $\mathbf{G}^y$, forms the full sampling grid $\mathbf{G}$ for splatting and slicing.

\vspace{1ex}
\noindent\textbf{Blurring.}
Traditional bilateral grids use fixed kernels \cite{chen2007real} or energy functions \cite{barron2016fast,chen2016bilateral} for operations like Gaussian denoising \cite{paris2006fast} and Laplacian enhancement \cite{paris2015local}, but this can be limiting. 
For example, reshaping a tensor $\mathbf{U} \in \mathbb{R}^{C\times H \times W}$ with $C=C_{\Gamma}\times R$ into a bilateral grid $\mathbf{\Gamma} \in \mathbb{R}^{C_{\Gamma}\times H \times W \times R}$ can inadequately capture the range domain.
Our method projects feature maps onto a sparse, high-dimensional grid as per Eq. \eqref{eq:splatting}, and uses learnable convolutional layers for adaptive filtering.

\vspace{1ex}
\noindent\textbf{Reducer.}
The reducer reduces channel size while doubling the spatial dimensions of the feature map. 
In transformers, it's a patch expanding module, mirroring Swin transformers' patch merging \cite{liu2021swin}. 
For ConvNets, it functions as a transpose convolution.

\subsection{Displacement Field Manipulation}
\label{sec:dis_manipulate}
We briefly review deformable image registration (DIR), followed by derivation of displacement field manipulation via a bilateral grid. 
DIR aligns a moving image $\mathbf{I}_m$ with a fixed image $\mathbf{I}_f$ using spatial mapping $\mathbf{\phi}(x) = x + \mathbf{u}(x)$, where $\mathbf{u}(x)$ is the displacement at $x$ in domain $\Omega \subset \mathbb{R}^{H\times W\times D}$. 
This warps $\mathbf{I}_m$ for voxel correspondence with $\mathbf{I}_f$, using linear interpolation for non-grid positions.
Unsupervised learning estimates deformation field $\phi$ through a network $F_{\theta}$, optimizing weights $\theta$ via a composite loss function $\mathcal{L}$. This combines dissimilarity between $\mathbf{I}_m$ and $\mathbf{I}_f$, and deformation field smoothness:
\begin{equation}
    \mathcal{L} = \mathcal{L}_{sim}(\mathbf{I}_f,\mathbf{I}_m \circ \phi) + \mathcal{L}_{sim}(\mathbf{G}_f^r,\mathbf{G}_m^r \circ \phi) + \lambda \mathcal{L}_{reg}(\phi),
    \label{eq:loss_general}
\end{equation}
where $\mathbf{G}_f^r$ and $\mathbf{G}_m^r$ are derived from $\mathbf{I}_f$ and  $\mathbf{I}_m$ via guidance mapping.
For cardiac registration, we use the loss function defined in Eq.~\eqref{eq:loss_general}. 
In lung registration, this is augmented with an additional Dice loss using segmentation masks and target registration loss based on keypoints.

\vspace{1ex}
\noindent\textbf{Displacement Field Manipulation.}
Using a bilateral grid, we can create a complete displacement field solely from keypoints. 
Given moving keypoint $\mathbf{p}_m \in \mathbb{R}^{N\times 3}$ and corresponding fixed keypoints $\mathbf{p}_f \in \mathbb{R}^{N\times 3}$, a sparse displacement field is formed by setting $ \mathbf{u}[\mathbf{p}_f(i)]=\mathbf{p}_m(i)-\mathbf{p}_f(i)$ for each $i \in \{1,2,...,N\}$, where $\mathbf{p}_*(i)$ denotes the $i_{th}$ point in the set. 
For all other points $\mathbf{u}(x)$ is a zero vector.
Given the guidance map $\mathbf{G}^{r}$ derived from either a guidance map generator or the raw image, the sparse displacement field $\mathbf{u}$ can be projected onto a bilateral grid as $\mathbf{\Gamma}=\mathcal{D}(\mathbf{u},\mathbf{G};\mathcal{K})$. 
Here, $\mathbf{G}$ includes mesh grids and $\mathbf{G}^{r}$, and $\mathcal{K}$ is a linear sampling kernel.
The task then is to minimize the equation $\argmin_{\tilde{\mathbf{\Gamma}}} \sum{||\text{grad}(\tilde{\mathbf{\Gamma}})||^2}$, subject to the constraint $\tilde{\mathbf{\Gamma}}(x)=\mathbf{\Gamma}(x)$, for every $x$ that $\mathbf{\Gamma}(x)\neq \mathbf{0}$.
This is used to fill up the zero values in the grid.
The optimization outlined in the equation can be efficiently executed through convolution and seamlessly integrated into neural network training.

%% file: docs/results.tex
\begin{table}[t!]
\centering
\resizebox{1.0\columnwidth}{!}{
\begin{tabular}{lcccccc}
\hline
\hline
Model   & Avg. ($\%$)  & RV ($\%$) & LVM ($\%$) & LVBP ($\%$) & HD95 (mm) $\downarrow$ &SDlogJ $\downarrow$ \\ \hline
Initial & 58.14 & 64.50 & 48.33 & 61.60 & 11.95 & - \\
\hline
ANTs  & 71.04 & 68.61 & 67.53 & 76.96 & 13.15 & 0.056\\
Demons  & 72.37 & 70.85 & 69.34 & 76.93 & 11.46 &  0.031\\
Bspline  & 74.36 & 72.18 & 71.68 & 79.22 & 11.18 &  \textbf{0.030}\\
\hline
VoxelMorph	& 76.35 & 74.69 & 73.19 & 81.15 & 9.28 & 0.049 \\
TransMorph & 76.89 & 75.39 & 73.52 & 81.75 & 9.11 & 0.049 \\
FourierNet & 77.04 & 75.30 & 73.88 & 81.96 & 9.10 & 0.045 \\
LKU-Net & 77.10 & 75.16	& 74.20 & 81.75 & 9.14 & 0.048 \\
DeBG	& 77.36 & 76.05 & 74.41 & 81.61 & 8.75 & 0.042 \\
\hline
Res-Slicer & \textbf{79.20} & \textbf{78.14} & \textbf{76.31} & \textbf{83.15} & \textbf{8.33} & 0.050 \\
\hline
\hline
\end{tabular}
}
\caption{
Quantitative evaluation of different models on the ACDC dataset, highlighting top scores in bold. 
Metrics include Average Dice (\%), RV Dice (\%), LVM Dice (\%), LVBP Dice (\%), HD95 (mm), and SDlogJ, with a $\downarrow$ indicating that lower values are better.
}
\label{tab:acdc}
\end{table}

\section{Applications, Experiments \& Results}
We evaluate the proposed Slicer Network across three medical image analysis applications: unsupervised cardiac registration using cine-MRI, keypoints-based lung registration between inspiration and expiration phases in CT scans, and skin lesion segmentation through dermoscopy. 
Common implementation aspects are outlined here, with individual application details discussed in their respective sections.

\vspace{1ex}
\noindent\textbf{General Implementation Details.}
Experiments are performed in Python 3.7 using PyTorch 1.9.0 \cite{paszke2019pytorch}. 
Training takes place on a machine with an A100 GPU, 32GB of memory, and a 16-core CPU. 
Splatting is based on CUDA code from \cite{zhang2023deda}, compiled with CUDA 11.1. 
For slicing, PyTorch's \emph{grid\_sample()} function is used.
Hyperparameters are optimized using grid search for all methods.



\subsection{Cardiac Cine-MRI Registration}
The ACDC dataset, with 150 subjects (100 for training, 50 for testing), is used for cardiac registration. 
Each subject has End-diastole (ED) and End-systole (ES) phase images with segmentation masks.
The intra-subject registration requires aligning images from both ED to ES and ES to ED, yielding 300 training pairs ($[100+50] \times 2$). 
We use 170 pairs for training, 30 for validation, and the remaining 100 pairs for testing. 
All images are resampled to $1.8\times1.8\times10$ mm spacing and processed to a size of $128\times128\times16$.

\vspace{1ex}
\noindent\textbf{Why Cardiac Registration.}
Cardiac registration between ED$\rightarrow$ES and ES$\rightarrow$ED presents a unique challenge due to its asymmetrical nature. 
Key objects like the left ventricular myocardium (LVM), left ventricle blood pool (LVBP), and right ventricle (RV) show smooth internal contrasts but differ across edges. 
This aligns with the Slicer Network's assumption, as the displacement fields required are also \textit{piece-wise smooth}.

\vspace{1ex}
\noindent\textbf{Baseline Methods.}
We benchmark the Slicer Network against leading registration models including VoxelMorph \cite{balakrishnan2019voxelmorph}, TransMorph \cite{chen2022transmorph}, LKU-Net \cite{jia2022u}, and Fourier-Net \cite{jia2023fourier}. 
Traditional methods ANTs \cite{avants2011reproducible}, Demons  \cite{vercauteren2007diffeomorphic}, and Bspline \cite{marstal2016simpleelastix} are also included.
Additionally, we include an implementation of the deep bilateral grid (denoted as DeBG), previously utilized in image manipulation \cite{gharbi2017deep} and stereo matching \cite{xu2021bilateral}, as another comparative baseline.

\vspace{1ex}
\noindent\textbf{Implementation Details.}
Our network, optimized using the Adam optimizer with a learning rate of 1e-4, a batch size of 4, and cosine decay, trains for 500 epochs. 
We employ Mean Square Error (MSE) for similarity loss and an L2 norm on deformation field spatial gradients as a smoothness regularizer ($\lambda=0.01$), following \cite{dalca2019unsupervised,balakrishnan2019voxelmorph}, with seven integration steps in the diffeomorphic layer. 
All deep learning models follow these settings.
For our method, named Res-Slicer, we use a ResNet-like encoder. 
The guidance map is sourced from the fixed image's feature map, with MSE loss applied between the fixed and moving images post-guidance convolution. 
For DeBG, aside from the encoder and splatting adapted from \cite{ghafoorian2017deep,xu2021bilateral}, all other aspects align with our Res-Slicer setup.
For DeBG, FourierNet, and Res-Slicer, downsampling is omitted in the axial direction due to thickness considerations.

\begin{figure}[!t]
    \centering
    \includegraphics[width=0.9\columnwidth]{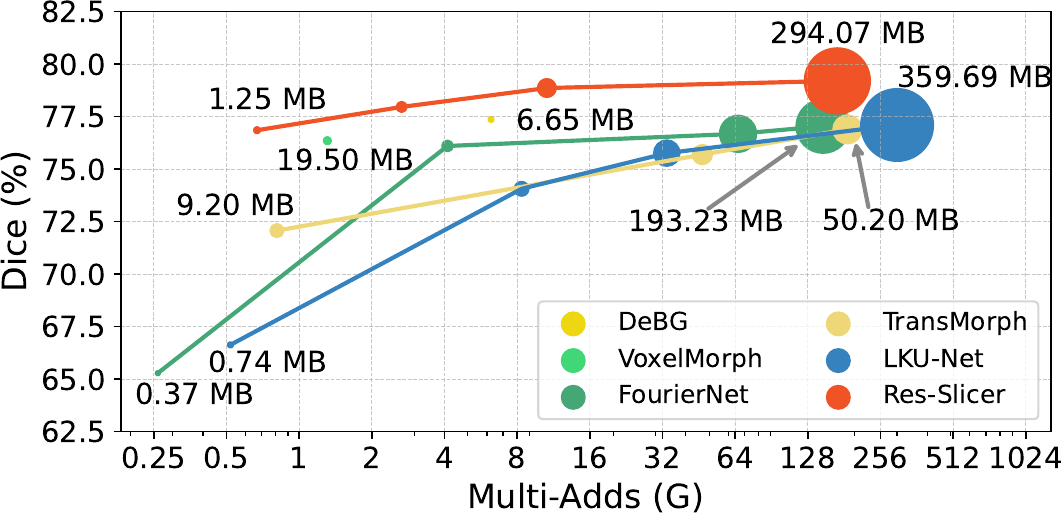}
    \caption{
    Trade-off between average Dice (\%) and computational complexity for the ACDC dataset, comparing parameter size and multi-add operations (in G) on a log-scaled x-axis. 
    TransMorph is shown in three complexities: tiny, small, and normal. 
    Complexity for FourierNet, LKU-Net, and Res-Slicer is adjusted by varying the network's initial channel count.
    }
    \label{fig:acdc_complexity}
\end{figure}

\vspace{1ex}
\noindent\textbf{Evaluation Metrics.}
Aligned with standard practices \cite{balakrishnan2019voxelmorph,chen2022transmorph}, our evaluation uses the Dice Similarity Coefficient (Dice) and the 95th percentile Hausdorff Distance (HD95) for anatomical alignment. 
We evaluate diffeomorphism quality using the standard deviation of the Jacobian determinant's logarithm (SDlogJ). 
Additionally, network complexity is assessed by measuring parameter size and multi-add operations.

\vspace{1ex}
\noindent\textbf{Registration Accuracy.}
Table \ref{tab:acdc} presents a comparison of methods, showing all produced smooth displacement fields with low SDlogJ. 
Learning-based methods generally surpass traditional ones in accuracy. 
Res-Slicer leads in all accuracy metrics, with DeBG, using shuffled channels for range dimension, outperforming others but not matching Res-Slicer. 
FourierNet competes well, achieving accuracy comparable to other methods but with reduced complexity.
This suggests that increasing the \textit{effective receptive field (ERF)} through strategic upsampling from deeper feature maps is beneficial for cardiac registration.
Results from Res-Slicer and DeBG emphasizes the importance of maintaining edge detail.

\vspace{1ex}
\noindent\textbf{Complexity Analysis.}
We evaluate network complexity using parameter size (MB) and multi-add operations (G). 
Complexity is adjusted based on the parameter size for each network. 
Fig. \ref{fig:acdc_complexity} compares the accuracy and complexity of Res-Slicer against other methods. 
Res-Slicer shows an optimal trade-off; with comparable Dice (76.85\%), it reduces parameter size and multi-add operations by 81.0\% and 89.0\% relative to DeBG, and by 99.7\% and 99.8\% compared to LKU-Net.

\begin{table}[t!]
\centering
\resizebox{1.0\columnwidth}{!}{
\begin{tabular}{lcccc}
\hline
\hline
Model   & Lung ($\%$) & TRE (mm)  $\downarrow$ & HD95 (mm) $\downarrow$ &SDlogJ $\downarrow$ \\ \hline
Initial & 79.53 & 14.64 & 206.45 & -  \\
\hline
TransMorph  & 93.25 & 10.23  & 160.07   & \textbf{0.143} \\
VoxelMorph  & 93.44 & 10.37  & 53.61 & 0.197 \\
FourierNet  & 93.98 & 8.74 & 18.36 & 0.205 \\
LKU-Net     & 94.91 & 8.44 & \textbf{12.30} & 0.218 \\
\hline
Zero-Shot (KPS)  & 83.36 & \textbf{2.35} & 186.52 & 0.304 \\
Point-Slicer& \textbf{95.77} & 2.66 & 16.431 & 0.221 \\
\hline
Zero-Shot (LMS)  & 80.21 & 0.20 & 207.50 & 0.220 \\
\hline
\hline
\end{tabular}
}
 \vspace{-1ex}
\caption{
Quantitative evaluation of different models on the Lung CT dataset, highlighting top scores in bold. 
Metrics include Average Dice of lung region (\%), TRE (mm), HD95 (mm), and SDlogJ, with a $\downarrow$ indicating that lower values are better.
}
\label{tab:lung_ct}
\end{table}

\subsection{Keypoints-Based Lung CT Registration}
We use a lung CT dataset \cite{hering10learn2reg} for keypoints-based image registration, featuring 20 subjects with paired expiration and inspiration phase scans. 
Each scan includes auto-generated keypoints with corresponding matches in both phases. 
For testing, manual landmarks are available for three subjects, with the remaining 17 used for training. 
CT scans, sized $192 \times 192 \times 208$, are preprocessed to a voxel spacing of $1.75 \times 1.25 \times 1.75$ mm.

\vspace{1ex}
\noindent\textbf{Why Lung CT Registration.}
Lung CT registration between expiration and inspiration phases present large and complex lung deformations. 
This process requires accurate alignment of both expansive anatomical regions and specific landmarks.
Existing models often struggle with the sparse keypoints in unsupervised registration. 
The Slicer Network uniquely addresses this by offering zero-shot learning from keypoints (as elaborated in Sec. \ref{sec:dis_manipulate}) for generating deformation fields, seamlessly incorporating these elements into its training process.

\vspace{1ex}
\noindent\textbf{Baseline Methods \& Evaluation Metrics.}
For lung CT registration, we maintain the same learning-based baseline methods used in cardiac registration, excluding DeBG due to its inability to handle keypoints with channel-based range representation. 
Alongside the metrics previously employed, we introduce the Target Registration Error of landmarks (TRE) to our evaluation criteria.

\begin{figure}[b!]
    \centering
     \vspace{-1ex}
    \includegraphics[width=0.95\columnwidth]{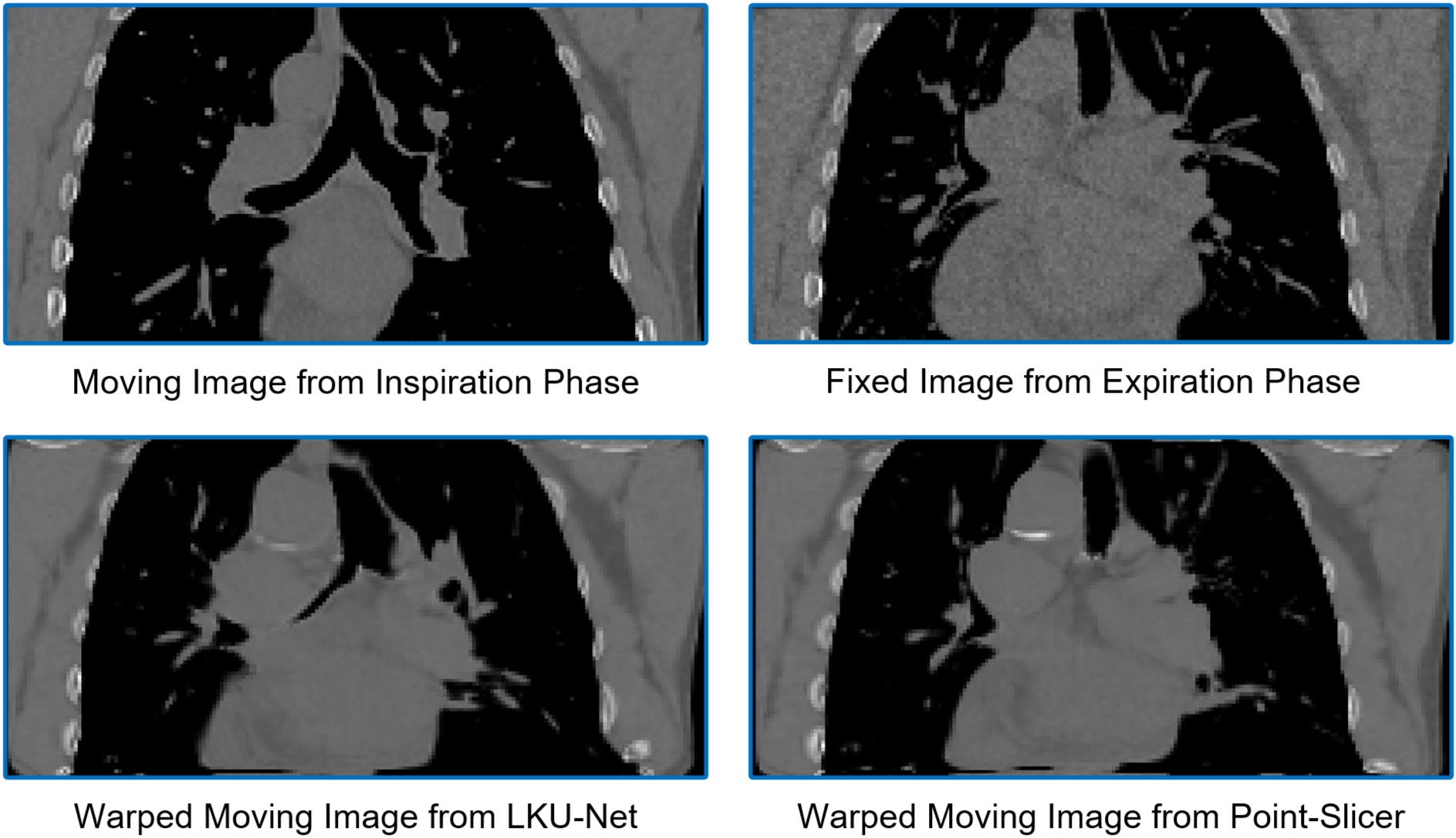}
    \caption{
    Visual comparison of lung CT dataset registration quality between LKU-Net and our Point-Slicer. 
    The inspiration scan and its warped counterpart are cropped to match the expiration scan's size due to the lungs' partial visibility.
    }
    \label{fig:lung_ct_reg_examples}
\end{figure}

\vspace{1ex}
\noindent\textbf{Implementation Details.}
Training parallels cardiac registration, with modifications including a batch size of 1 and 200 total epochs. 
We incorporate Dice and TRE losss ($\lambda=0.1$).
Our method, named Point-Slicer, uses LKU-Net as the backbone.
The fixed image undergoes guidance mapping, followed by keypoints being projected onto a bilateral grid. 
This generates a dense displacement field (described in Sec. \ref{sec:dis_manipulate}), which is then used to warp the moving image prior to its processing through the backbone.
Zero-Shot (KPS) involves using the displacement field generated from raw fixed images and auto-generated keypoints (KPS) as the final outcome. 
Zero-Shot (LMS) uses ground-truth landmarks (LMS) as a reference and for analytical purposes only.
Both Zero-Shot learners apply the method outlined in Sec. \ref{sec:dis_manipulate}, relying on raw fixed images instead of learnable guidance mapping.

\begin{table}[t!]
\centering
\resizebox{1.0\columnwidth}{!}{
\begin{tabular}{lccccc}
\hline
\hline
Model & Dice (\%) & Prec (\%) & Sens (\%) & F1 Score (\%) & HD95 $\downarrow$  \\ \hline
Unet & 84.34 & 87.41 & 87.39 & \FPeval{\result}{round(2*87.41*87.39/(87.41+87.39),2)}\result & 23.66 \\
Att-Unet  & 85.84 & 85.77 & 90.75 & \FPeval{\result}{round(2*85.77*90.75/(85.77+90.75),2)}\result & 21.50 \\
SAM & 87.20 & 84.57 & 93.48 & \FPeval{\result}{round(2*84.57*93.48/(84.57+93.48),2)}\result & 17.74 \\
Polyp-PVT & 88.14 & 86.95 & 93.09 & \FPeval{\result}{round(2*86.95*93.09/(86.95+93.09),2)}\result & 17.05 \\
Swin-Unet & 88.35 & 89.10 & 91.00 & \FPeval{\result}{round(2*89.10*91.00/(89.10+91.00),2)}\result & 17.95 \\
PVT-Cascade & 88.51 & 87.23 & 92.65 & \FPeval{\result}{round(2*87.23*92.65/(87.23+92.65),2)}\result & 16.55 \\
Tran-Unet & 88.93 & 86.76 & \textbf{94.40} & \FPeval{\result}{round(2*86.76*94.40/(86.76+94.40),2)}\result & 16.60 \\
Swin-Slicer & \textbf{90.02} & \textbf{90.57 }& 92.12 & \textbf{\FPeval{\result}{round(2*90.57*92.12/(90.57+92.12),2)}\result} & \textbf{15.16} \\ 
\hline
\end{tabular}
}
 \vspace{-1ex}
\caption{
Quantitative evaluation of different models on the test set of ISIC2018 dataset, highlighting top scores in bold. 
Metrics include Average Dice (\%), Prec (\%), Sens (\%), F1 Score (\%) and HD95 (in pixels), with a down arrow indicating that lower values are better.
}
\label{tab:isic2018}
\end{table}

\vspace{1ex}
\noindent\textbf{Registration Accuracy.}
Table \ref{tab:lung_ct} compares various methods, including Zero-Shot learners. 
Zero-Shot (KPS) excels in TRE but lags in Dice and HD95, due to its well-aligned lung landmarks but sparse keypoints leading to inadequate lung boundary alignment. 
Point-Slicer, utilizing both keypoints and lung mask data, attains a well-balanced Dice and TRE, leading to improved anatomical region alignment, as depicted in Fig. \ref{fig:lung_ct_reg_examples}, outperforming other techniques.
Zero-Shot (LMS), using landmark-generated fields, shows low TRE, indicating that improved keypoint descriptors could enhance landmark alignment performance. 
This also points to Point-Slicer's potential for end-to-end joint training of keypoint descriptor and anatomical region alignment, a direction for future research.

\subsection{Skin Lesion Segmentation}
\label{sec:skin_lesion_results}
We evaluate the Slicer Network for skin lesion segmentation using the ISIC2018 dataset \cite{codella2019skin,tschandl2018ham10000}, comprising 2594 training, 100 validation, and 1000 testing images. 
Adhering to the official dataset divisions for training, validation, and testing, as specified on the dataset's website\footnote{\url{https://challenge.isic-archive.com/landing/2018}}, we ensure consistency and repeatability without manual data partitioning.

\vspace{1ex}
\noindent\textbf{Why Skin Lesion Segmentation.}
Dermoscopic skin lesion segmentation presents a unique challenge due to the variability in lesion shapes, sizes, and textures, despite consistent internal textures. 
This scenario aligns well with the Slicer Network's design assumpation, particularly as the resulting segmentation masks are \textit{piece-wise constant}.

\vspace{1ex}
\noindent\textbf{Baseline Methods.}
We evaluate the Slicer Network against leading models including SAM \cite{Kirillov_2023_ICCV}, TransUnet \cite{chen2021transunet}, Swin-Unet \cite{cao2022swin}, Polyp-PVT \cite{dong2023PolypPVT}, PVT-Cascade \cite{rahman2023medical}, Unet \cite{ronneberger2015u}, and Att-Unet \cite{oktay2018attention}, fine-tuning the SAM backbone without additional prompts.

\vspace{1ex}
\noindent\textbf{Implementation Details and Evaluation Metrics.}
Training employs the Adam optimizer \cite{kingma2014adam} with an initial learning rate of 1e-4 and a multi-step scheduler reducing it by half at 50\%, 70\%, and 90\% of the 70 total epochs. 
Images are resized to $(224,224)$, with batches of 16. 
Segmentation is evaluated using Dice for overlap, Sensitivity (Sens), Precision (Prec), and F1-score (the harmonic mean) for detection accuracy, along with HD95 (in pixels) for boundary precision.
Our network, termed Swin-Slicer, incorporates the Swin transformer's tiny version \cite{liu2021swin} as encoder.

\vspace{1ex}
\noindent\textbf{Segmentation Accuracy.}
Table \ref{tab:isic2018} shows Swin-Slicer outperforming transformer-based and traditional ConvNet methods on the ISIC2018 dataset in Dice, F1-Score, and HD95. 
This demonstrates superior segmentation precision and boundary adherence. 
While Trans-Unet shows highest sensitivity, it lags in precision. 
In contrast, Swin-Slicer achieves a balanced trade-off, leading in F1-Score without sacrificing sensitivity.

\begin{figure}[!t]
    \centering
    \includegraphics[width=1.0\columnwidth]{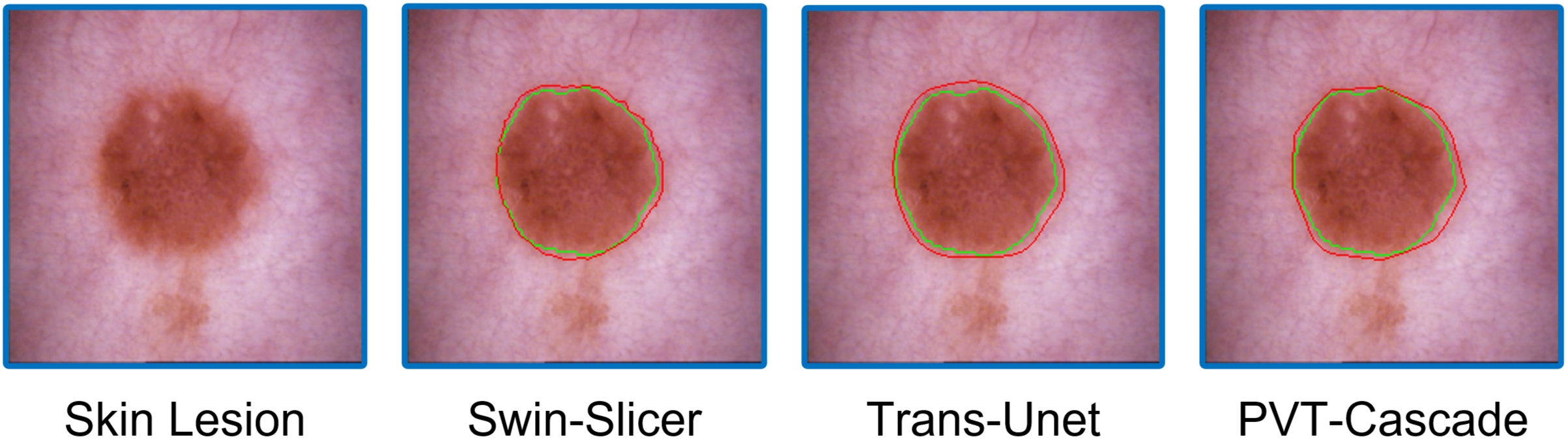}
    \caption{
        Qualitative segmentation results from the top-performing methods based on Dice scores: Swin-Slicer, Trans-Unet, and PVT-Cascade. 
        The green contour represents the ground truth delineation of the lesion, and the red contour indicates the model's prediction.
    }
    \label{fig:isic2018_examples}
\end{figure}

\vspace{1ex}
\noindent\textbf{Qualitative Results.}
Swin-Slicer effectively uses just a $7\times7$ feature map from the Swin transformer encoder for the slicer, yet, as Fig. \ref{fig:isic2018_examples} shows, it achieves superior delineation of lesion boundary. 
This success is attributed to maintaining the ERF by slicing directly from the lifted bilateral grid, coupled with the guidance map from raw images preserving lesion boundary details.

\subsection{Discussions}
\noindent\textbf{Effects of Sampling Rate $s_r$.}
The sampling rate $s_r$ affects the range domain ($\mathcal{R}$). 
At $s_r=1$, the range dimension simplifies to 1, transforming the process into a gated attention where the attention map, derived from the guidance map, gates the linearly upsampled encoder output. 
Decreasing $s_r$ increase the dimensionality in range domain and enhances edge distinction in spatial domain but also increases computational demands. 
Experimentally, reducing $s_r$ from $s_r=1$ to $s_r=1/64$ reveals that accuracy in cardiac, lung registration, and lesion segmentation improves as $s_r$ decreases, reaching optimal points at $s_r=1/8$, $s_r=1/8$, and $s_r=1/32$, respectively, before declining.
The smaller optimal $s_r$ for lesion segmentation reflects its greater need for precise boundary handling.

\vspace{1ex}
\noindent\textbf{Effects of Sampling Rate $s_s$.}
The sampling rate $s_s$ influences the spatial domain ($\mathcal{S}$). 
At $s_s=1$, spatial dimensions match the original image. 
Increasing $s_s$ results in smoother outcomes, larger ERF in neural networks, and reduced computational load. 
Increasing $s_s$ from $s_s=1$ to $s_s=32$ improves accuracy in cardiac and lung registration, and lesion segmentation, peaking at $s_s=8$ for registrations and $s_s=32$ for segmentation, then decreasing. 
The higher optimal $s_s=32$ for lesion segmentation is consistent with the spectrum analysis below.

\begin{figure}[!t]
    \centering
    \includegraphics[width=1.0\columnwidth]{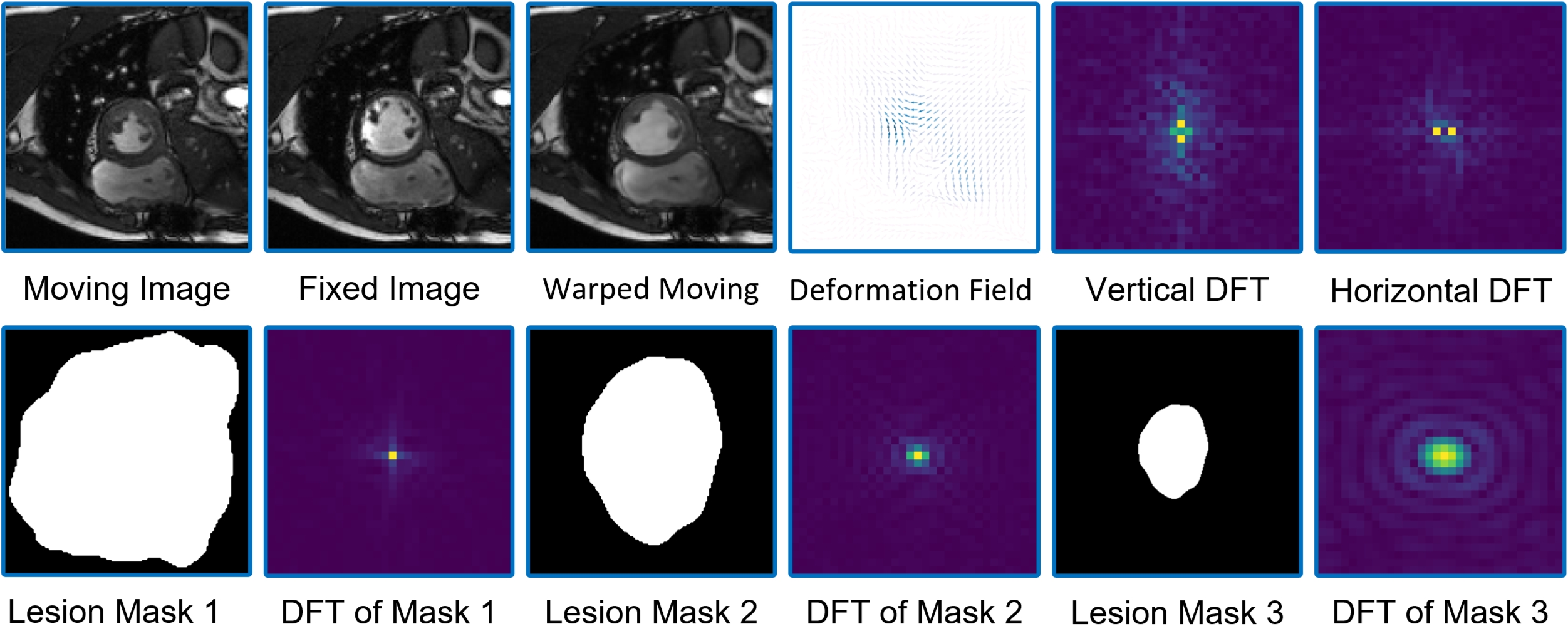}
    \caption{
    The first row shows cardiac registration visualizations with Res-Slicer, displaying $32 \times 32$ resized 2D deformation fields and their Discrete Fourier Transforms (DFTs) for clarity. 
    The second row features lesion masks and their DFTs, with lesion mask 3 being a smaller scaled version of mask 2. 
    DFT magnitudes are normalized to (0,1) and resized to $32 \times 32$ for enhanced visualization quality.
    }
    \label{fig:visual_examples}
\end{figure}

\vspace{1ex}
\noindent\textbf{Spectrum Analysis.}
Fig. \ref{fig:visual_examples} shows the outcome deformation field from cardiac registration and binary masks from skin lesion segmentation, along with their magnitude spectrums. 
Larger masks tend to concentrate more in the low-frequency range. 
Additionally, the deformation field exhibits more higher-frequency components compared to the lesion mask, reflecting the varying spatial sampling rates needed for image registration versus lesion segmentation.
While spectrum analysis has previously been applied in image registration \cite{jia2023fourier,wang2020deepflash} and segmentation \cite{zhou2023xnet} and proved effective for certain applications, our work is unique in introducing a unified framework aimed at enhancing a broad range of medical applications.

\vspace{1ex}
\noindent\textbf{Limitations.}
The development of the Slicer Network has revealed three key limitations. 
First, its suitability is limited for applications demanding high-frequency components in the output, such as curvilinear structure segmentation and small targets detection. 
Second, while the bilateral grid is effective, more efficient high-dimensional filtering methods like adaptive manifolds \cite{gastal2012adaptive} could offer improvements. 
Finally, the guidance map generation, essentially a metric learning process, remains an area for further exploration and development.

%% file: docs/conclusion.tex
\section{Conclusions}
\label{sec:conclusion}
Our paper introduces the Slicer Network, a novel neural network architecture for medical image analysis, effectively addressing two questions from Sec. \ref{sec:intro}. 
For \textbf{Q1}, the network incorporates a learnable cross-bilateral grid with guidance mapping, enhancing the \textit{effective receptive field} while preserving boundary details. 
For \textbf{Q2}, in scenarios where both input images and outcomes are \textit{piece-wise smooth}, the Slicer Network offers reduced computational demands without sacrificing accuracy. 
Its proven effectiveness across three medical applications indicates potential advantages for other medical tasks that meet the same underlying assumption.